\documentstyle[aps,floats]{revtex}
\begin{document}
\draft
\twocolumn[\hsize\textwidth\columnwidth\hsize\csname 
           @twocolumnfalse\endcsname
\title{Gravitational radiation from a particle
       in circular orbit around a black hole. \\
       VI. Accuracy of the post-Newtonian expansion}
\author{Eric Poisson$^*$}
\address{McDonnell Center for the Space Sciences, 
         Department of Physics, Washington University,
         St.~Louis, Missouri 63130}
\maketitle
\begin{abstract}
\widetext
A particle of mass $\mu$ moves on a circular orbit around
a nonrotating black hole of mass $M$. Under the assumption
$\mu \ll M$ the gravitational waves emitted by such a binary
system can be calculated exactly numerically using black-hole 
perturbation theory. If, further, the particle is slowly moving, 
$v= (M\Omega)^{1/3} \ll 1$ (where $v$ and $\Omega$ are respectively 
the linear and angular velocities in units such
that $G=c=1$), then the waves can be calculated approximately 
analytically, and expressed in the form of a post-Newtonian 
expansion. We determine the accuracy of this expansion in a 
quantitative way by calculating the reduction in 
signal-to-noise ratio incurred when matched filtering the exact 
signal with a nonoptimal, post-Newtonian filter. We find that 
the reduction is quite severe, approximately 25\%, for 
systems of a few solar masses, even with a post-Newtonian 
expansion accurate to fourth order --- $O(v^8)$ --- beyond 
the quadrupole approximation. Most of this reduction is 
caused by post-Newtonian theory's inability to correctly 
locate the innermost stable circular orbit, which here is at $r=6M$ 
(in Schwarzschild coordinates). Correcting for this yields 
reductions of only a few percent. 
\end{abstract}
\pacs{PACS numbers: 04.25.Nx, 04.30.-w, 97.60.Jd, 97.60.Lf}
\vskip 2pc]
\narrowtext

Inspiraling compact binary systems, composed of neutron
stars and/or black holes, have been identified 
\cite{Thorne1987,Schutz} as the most 
promising source of gravitational waves for interferometric 
detectors such as LIGO (Laser Interferometer Gravitational-wave 
Observatory \cite{LIGO}) and VIRGO \cite{VIRGO}. 
These systems evolve under radiation
reaction, so that the gravitational-wave signal increases in 
amplitude as the frequency sweeps through the detector bandwidth, 
from approximately 10 Hz to 1000 Hz. 

Extraction of the information contained in the gravitational
waves, most notably about the masses and spins of the companions, 
will necessitate the construction of accurate model signals, 
known as templates 
\cite{Finn,FinnChernoff,CutlerFlanagan,PoissonWill}. 
The extraction makes use of the well-known 
technique of matched filtering \cite{WainsteinZubakov,Helstrom}, 
in which the signal is integrated 
against a choice of templates in order to identify the true value 
of the source parameters. It has been established that these 
templates will need to reproduce the signal's {\it phasing} 
especially accurately \cite{FinnChernoff,CutlerFlanagan,Cutleretal}. 
This is because the signal undergoes a
number of oscillations approximately equal to 16 000 (if the 
system is that of two neutron stars) as the frequency sweeps 
through the detector bandwidth. If the template loses phase 
with respect to the true signal, even by so little as one cycle, 
then the signal-to-noise ratio is severely reduced, and the
information-extraction strategy severely impeded. Thus an 
accuracy of at least one part in 16 000 is required for the 
template's phasing.

The construction of accurate templates is currently the goal
of many gravitational-wave theorists \cite{Will}. 
It is clear that the
calculation must be based on some approximation to the 
equations of general relativity; the exact integration
of the equations governing the evolution of compact binary
systems is not currently amenable to numerical methods.  
The favored approach is post-Newtonian theory, which is 
based upon an assumption of slow motion: If $v$ denotes the 
orbital velocity, $M$ the total mass, and $r$ the orbital 
separation, then it is assumed that the orbital evolution 
is such that $v \sim (M/r)^{1/2} \ll 1$. We set $c=G=1$. 

To date, templates have been calculated accurately through
second post-Newtonian order \cite{BDIWW,BDI,WW}, 
that is, order $v^4$ beyond
the leading-order, quadrupole-formula expression. The question 
considered in this paper is whether this calculation is accurate 
enough for the purpose of information extraction. We shall see,
as other authors have argued 
\cite{paperII,TagoshiNakamura,TagoshiSasaki}, 
that the answer is negative. The
post-Newtonian expansion converges extremely slowly, if at all.
Calculations must be pushed to a much higher order.

The exact treatment of the orbital evolution of a compact binary 
system whose companions have comparable masses is currently beyond 
reach. The question posed in the preceding paragraph cannot, 
therefore, be answered in this context. However, if instead 
we consider a system composed of a particle with small 
mass orbiting a black hole with much larger mass, then 
the question {\it can} satisfactorily be answered. This limiting 
case indeed allows for an exact treatment, which can then be compared 
with an approximate, slow-motion treatment. The degree of validity
of the approximation can thereby be ascertained. Our hope is 
that our conclusions, based upon the small-mass-ratio limit, will 
remain qualitatively valid for systems with large mass ratios.

The gravitational waves generated by a particle in circular
orbit around a black hole have been the topic of this series 
of papers \cite{paperII,paperI,paperIII,paperIV,paperV}. 
If $\mu$ denotes the particle mass, and $M$ that
of the black hole (which is here assumed to be nonrotating),
then the assumption $\mu/M \ll 1$ ensures that the gravitational 
perturbations produced by the orbiting particle are small and 
governed by Teukolsky's linear wave equation 
\cite{Teukolsky}. This equation can 
be integrated using straightforward numerical methods, and 
the gravitational waveform is thus determined exactly.

We will focus mainly on the phasing of the waves, which is
determined by $df/dt$, the rate at which the 
gravitational-wave frequency changes with time. For
circular orbits $f=\Omega/\pi$, where $\Omega = d\phi/dt$
is the angular velocity, and $df/dt$ can be expressed as
\begin{equation}     
\frac{df}{dt} = \frac{dE/dt}{dE/df}.
\label{1}
\end{equation}
Here, $dE/dt$ is the rate at which gravitational waves
remove energy from the system, and $dE/df$ expresses
the relation between orbital energy and wave frequency.

The solid curve in Fig.~1 is a plot of 
\begin{equation}
P(v) \equiv \frac{dE/dt}{(dE/dt)_N},
\label{2}
\end{equation}
where 
\begin{equation}
v = (\pi M f)^{1/3}
\label{3}
\end{equation}
is the orbital velocity and $(dE/dt)_N = 
- 32\mu^2 v^{10} / 5 M^2$ the Newtonian, 
quadrupole-formula expression for
the gravitational-wave luminosity. The plot was obtained by 
numerically integrating the Teukolsky
equation, along the lines described in Ref.~\cite{paperII}. 
An expression for $dE/df$ can be obtained by integrating
the geodesic equations for circular orbits. The orbital 
energy is defined as $E=-p_t$, were $p^\alpha$ is the particle's 
four momentum; it is a constant of the motion in the absence of 
radiation reaction. We obtain
\begin{equation}
Q(v) \equiv \frac{dE/df}{(dE/df)_N} =
(1-6v^2)(1-3v^2)^{-3/2},
\label{4}
\end{equation}
where $(dE/df)_N = - \pi \mu M / 3v$ is the
Newtonian expression. Equation (\ref{4}) and
the solid curve in Fig.~1 give through 
Eq.~(\ref{1}) an {\it exact} 
representation of $df/dt$.

To produce the solid curve in Fig.~1 
the Teukolsky equation had to be
integrated numerically. If, however, the 
small-mass-ratio approximation is combined with
a slow-motion approximation ($v\ll 1$), then the
Teukolsky equation can be integrated analytically
\cite{TagoshiSasaki,paperI,paperV,Sasaki}.
The resulting approximate result for $P(v)$ is 
\begin{eqnarray}
P(v) &=& 1 - 3.711 v^2 + 12.56 v^3 - 4.928 v^4 
- 38.29 v^5 
\nonumber \\ & & \mbox{}
+ (115.7 - 16.30 \ln v) v^6 
- 101.5 v^7 
\nonumber \\ & & \mbox{}
- (117.5 - 52.74 \ln v) v^8 + \cdots;
\label{5}
\end{eqnarray}
the various coefficients can be found in their analytic
form in Ref.~\cite{TagoshiSasaki}. Equation (\ref{5}) takes
the form of a post-Newtonian expansion for $P(v)$. The
first five terms, through $O(v^5)$, reproduce
in the small-mass-ratio limit 
the post-Newtonian calculation of 
Refs.~\cite{BDIWW,Blanchet}; the
remaining terms have not yet been calculated using
post-Newtonian theory. A post-Newtonian expansion
can also be given for $Q(v)$, by simply expanding
Eq.~(\ref{4}) about $v=0$. The result is
\begin{equation} 
Q(v) = 1 -
{\textstyle \frac{3}{2}} v^2 -
{\textstyle \frac{81}{8}} v^4 -
{\textstyle \frac{675}{16}} v^6 -
{\textstyle \frac{19845}{128}} v^8 
+ \cdots .
\label{6}
\end{equation}
Equations (\ref{5}) and (\ref{6}) give through 
Eq.~(\ref{1}) an {\it approximate} (post-Newtonian)
representation of $df/dt$.

\begin{figure}[t]
\special{hscale=55 vscale=55 hoffset=-10.0 voffset=-365.0
         angle=0 psfile=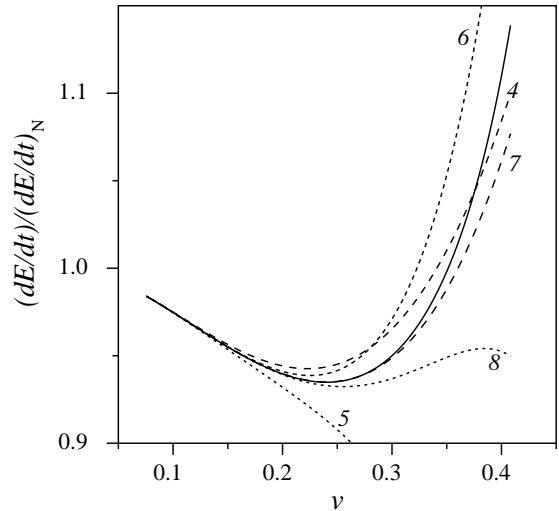}
\vspace*{3.0in}
\caption[Fig. 1]{Various representations of $(dE/dt)/(dE/dt)_N$
as a function of orbital velocity $v = (M/r)^{1/2} = 
(\pi M f)^{1/3}$. The solid curve represents the exact 
result $P(v)$, as calculated numerically. 
The various broken curves represent the
post-Newtonian approximations $P^{(n)}(v)$, for 
$n=\{4,5,6,7,8\}$. The smallest value of $v$ corresponds
to an orbital radius $r$ of $175M$; the largest value of 
$v$ corresponds to $r=6M$, the innermost stable circular orbit.}
\end{figure} 

Our goal in the sequel is to determine to which
extent the post-Newtonian representation of $df/dt$ 
reproduces the true phasing of the waves. We shall 
denote by $P(v)$ and $Q(v)$ the exact version of
these functions, while $P_n(v)$ 
and $Q_n(v)$ will denote  
the post-Newtonian representations 
truncated to the $n$th power of
$v$. For example, $P_4(v)$ represents the 
right-hand side of Eq.~(\ref{5}) with all terms 
of order $v^5$ and higher removed. 

Figure 1 displays the various curves $P_n(v)$ for
$n$ ranging from 4 to 8. It is seen that 
for $v>0.2$, these curves reproduce rather
poorly the exact curve $P(v)$. 
Moreover, the convergence of the post-Newtonian
expansion is also seen to be poor: adding a term
in the expansion does not necessarily make the
expression more accurate. Witness in particular
the poor quality of $P_5(v)$ compared with
that of $P_4(v)$. [Compare also $P_8(v)$
to $P_7(v)$.] We shall now attempt to determine,
in a quantitative manner, how much
of an obstacle this slow rate of convergence poses 
to information-extraction strategies. 

The starting point of our analysis is an expression for 
the reduction in signal-to-noise ratio incurred when 
the matched filtering of a gravitational-wave signal 
is carried out with a nonoptimal filter. If 
$h(t)$ denotes the true gravitational waveform, then 
optimal filtering is achieved by using $h(t)$ as a 
template. This is well-known to yield $S/N|_{\rm max}$, 
the largest possible value of the signal-to-noise 
ratio \cite{WainsteinZubakov,Helstrom}. 
If, instead, we adopt the post-Newtonian approximation 
$h_n(t)$ as a template, then filtering is not optimal, 
and $S/N|_{\rm actual}$, the actual value of the 
signal-to-noise ratio, is smaller than the maximum
possible value. The reduction in signal-to-noise ratio
can be calculated to be \cite{Apostolatos}
\begin{equation}
{\cal R}_n \equiv \frac{S/N|_{\rm actual}}{
S/N|_{\rm max}} = 
\frac{| (h | h_n) |}{
\sqrt{ (h|h) ( h_n| h_n ) }},
\label{7}
\end{equation}
where the inner product $(\cdot | \cdot)$ will be
defined presently.

We pause here to remark that in our calculations, 
we shall not allow the mass parameters $\mu$ and $M$ 
to take different values in $h(t)$ and 
$h_n(t)$. The problem considered here is 
{\it not} that of maximizing the signal-to-noise ratio 
over the source parameters in order to estimate their 
true value. We therefore do {\it not} address the 
issue of the systematic errors introduced in parameter 
estimation when using templates which are only approximations 
to the true general-relativistic signal 
\cite{CutlerFlanaganB}. Our considerations 
are less ambitious: the source parameters are assumed to 
be given, and the accuracy of the post-Newtonian templates
is quantitatively measured by ${\cal R}_n$, the relative
reduction in signal-to-noise ratio. This measure, we believe, 
is more satisfactory than the mere counting of the number of 
wave cycles contributed by various terms in the post-Newtonian 
expansion of the wave's phasing 
\cite{paperII,TagoshiNakamura,TagoshiSasaki}.

We now return to the definition of the inner product.
Let $\tilde{h}(f)$ and $\tilde{h}_n(f)$ denote,
respectively, the Fourier transforms of the functions
$h(t)$ and $h_n(t)$, with the convention 
$\tilde{g}(f) = \int g(t) e^{2\pi i f t}\, dt$ for
Fourier transforms. Let also $S(f)$ be the spectral
density of the detector noise, assumed to be a 
stationary, Gaussian random process. For detectors 
of the advanced-LIGO type it is appropriate to set
\cite{CutlerFlanagan}
\begin{equation}
S(f) = {\textstyle \frac{1}{5}} S_0 \Bigl[
(f_0/f)^4 + 2 + (f/f_0)^2 \Bigr]
\label{8}
\end{equation}
for $f > 10\ \mbox{Hz}$; for $f < 10\ \mbox{Hz}$ we
take $S(f) = \infty$. In Eq.~(\ref{8}), $S_0$ is
a normalization constant irrelevant for our purposes,
and $f_0$ is the frequency at which $S(f)$ is
minimum; we set $f_0 = 70\ \mbox{Hz}$. The inner
product introduced in Eq.~(\ref{7}) is then given
by \cite{CutlerFlanagan}
\begin{equation}
(g|h) = 2 \int_0^\infty \frac{
\tilde{g}^*(f) \tilde{h}(f) + 
\tilde{g}(f) \tilde{h}^*(f)}{S(f)}\, df,
\label{9}
\end{equation}
where an asterisk denotes complex conjugation. 

We must now specify $\tilde{h}(f)$ and $\tilde{h}_n(f)$, 
the gravitational waveforms in the frequency domain. 
We use an approximation \cite{CutlerFlanagan}
in which the {\it amplitude} of the waveform
is described accurately to Newtonian order, while its 
{\it phase} is described exactly in the case of 
$\tilde{h}(f)$, and by a post-Newtonian expansion in
the case of $\tilde{h}_n(f)$. Consequently, we have
\cite{CutlerFlanagan,PoissonWill}
\begin{eqnarray}
\tilde{h}(f) &=& {\cal A} f^{-7/6} 
\exp\bigl[ i \psi(f) \bigr], \nonumber \\
& & \label{10} \\
\tilde{h}_n(f) &=& {\cal A} f^{-7/6}
\exp\bigl[ i \psi_n(f) \bigr], \nonumber
\end{eqnarray}
where $\cal A$ is a constant. The phase
functions $\psi(f)$ and $\psi_n(f)$ are
constructed as follows.

We consider $\psi(f)$; $\psi_n(f)$ is dealt 
with similarly. Our starting point is the wave's
phasing in the time domain, which is determined
by $df/dt$ given above. Combining
Eqs.~(\ref{1})--(\ref{4}) we obtain $df/dt =
96\mu v^{11} P(v) / 5\pi M^3 Q(v)$.
Integration then yields
\begin{equation}
t(v)/M = t_i/M + \frac{5 M}{32 \mu}
\int_{v_i}^v \frac{Q(v')}{v^{\prime 9} P(v')}\, dv'
\label{11}
\end{equation}
for time as a function of velocity, and
\begin{equation}
\Phi(v) = \Phi_i + \frac{5M}{16\mu}
\int_{v_i}^v \frac{Q(v')}{v^{\prime 6} P(v')}\, dv'
\label{12}
\end{equation}
for the phase $\Phi = \int 2\pi f dt$. In 
Eqs.~(\ref{11}) and (\ref{12}), $v_i$ is an
arbitrary reference point, and $t_i = t(v_i)$,
$\Phi_i = \Phi(v_i)$. The frequency-domain
phase function $\psi(f)$ is obtained via the
stationary phase approximation 
\cite{CutlerFlanagan,PoissonWill,foot}, according to
which $\psi(f) = 2\pi f t(v) - \Phi(v) - \pi/4$. 
We therefore find
\begin{eqnarray}
\psi(f) &=& 2(t_i/M) v^3 - \Phi_i - \pi/4 
\nonumber \\ & & \mbox{} +
\frac{5M}{16\mu} \int_{v_i}^v 
\frac{(v^3-v^{\prime 3}) Q(v')}{v^{\prime 9} P(v')}\,
dv',
\label{13}
\end{eqnarray}
where, we recall, $v \equiv (\pi M f)^{1/3}$. 
Equations (\ref{10}) and (\ref{13}), together
with the fact that the gravitational-wave signal
must be cut off at a frequency $f_{\rm isco}$ 
corresponding to the innermost stable circular orbit, 
completely specify the Fourier transform of the 
waveform. We use $\pi M f_{\rm isco} =
(M/r_{\rm isco})^{3/2} = 6^{-3/2}$, where 
$r_{\rm isco}=6M$ is the radius of the innermost 
stable circular orbit of the Schwarzschild
spacetime. 

The calculation of ${\cal R}_n$ can now be
carried out. Straightforward manipulations, using
Eqs.~(\ref{7})--(\ref{10}), yield
\begin{equation}
{\cal R}_n = I^{-1} 
\int_{1/7}^{x_{\rm isco}} \frac{
x^{-7/3} \cos \Delta \psi}{
x^{-4} + 2 + 2x^2}\, dx,
\label{14}
\end{equation}
where $x=f/f_0$, $x_{\rm isco} = f_{\rm isco}/f_0$,
and 
\begin{equation}
\Delta \psi = \psi - \psi_n.
\label{15}
\end{equation}
The constant $I$ in Eq.~(\ref{14}) is given by
\begin{equation}
I = \int_{1/7}^{x_{\rm isco}} \frac{
x^{-7/3}}{x^{-4} + 2 + 2x^2}\, dx,
\label{16}
\end{equation}
and ensures that ${\cal R}_n = 1$ if $\psi(f)
= \psi_n(f)$. 

The numerical value of ${\cal R}_n$, for given 
$n$, depends on the value of the constants $t_i$ 
and $\Phi_i$ which appear in $\psi(f)$, and on the 
value of the constants $t_{ni}$ and $\Phi_{ni}$
which  appear in $\psi_n(f)$. To maximize 
${\cal R}_n$, we set $t_i = t_{ni}$
and choose $\Phi_i - \Phi_{ni}$ such that 
$\Delta \psi$ vanishes at the value of $f$ for which 
$x^{-7/3}(x^{-4} + 2 + 2 x^2)^{-1}$ is maximum. It is
easy to check that this occurs at $x_{\rm max} 
\simeq 0.6654$, so that $f_{\rm max} \simeq 46.58
\ \mbox{Hz}$. A simple calculation also shows that
$v_{\rm max} \equiv (\pi M f_{\rm max})^{1/3} 
\simeq 0.08966 (M/M_{\odot})^{1/3}$,
where $M_\odot$ denotes the mass of the Sun. With
these choices, $\Delta \psi$ becomes
\begin{equation}
\Delta \psi = \frac{5M}{16\mu} \int_{v_{\rm max}}^v
\frac{v^3-v^{\prime 3}}{v^{\prime 9}} 
\Biggl[ \frac{Q(v')}{P(v')} - 
\frac{Q_n(v')}{P_n(v')} \Biggr]\, dv'.
\label{17}
\end{equation}
It is a straightforward numerical problem \cite{NumRec} 
to compute $\Delta \psi$ for a given $v = (\pi M f)^{1/3}$ 
and to then evaluate the integral to the right-hand side of
Eq.~(\ref{14}). 

The calculation presented above is, strictly speaking, only 
applicable to binary systems with small mass ratios. In this 
limit, the only place in which $\mu/M$ appears is
as an overall multiplicative factor in Eq.~(\ref{17}). In
this limit, therefore, the phase lag $\Delta \psi$ scales
exactly as $M/\mu$. As the mass ratio is allowed to increase, 
our results for $P(v)$, $Q(v)$, and  
their post-Newtonian analogues, must be corrected \cite{foot}. 
For example, the constant coefficients in Eqs.~(\ref{5}) 
and (\ref{6}) would acquire $\mu/M$-dependent 
corrections \cite{BDIWW}. 
Our expressions for $(dE/dt)_N$ and $(dE/df)_N$, however, 
stay valid for large mass ratios, provided that 
$\mu$ is then interpreted as the system's {\it reduced} mass, 
and $M$ as the system's {\it total} mass. 

In the following we will let $\mu/M$ become large, without 
modifying our expressions for $P(v)$, $Q(v)$, and
their post-Newtonian analogues. We will {\it assume},
without justification, that the qualitative behavior of 
these functions is not appreciably affected by the 
finite-mass-ratio corrections [which are not known, apart 
from the terms of lowest order in $P_n(v)$ and 
$Q_n(v)$]. We will therefore apply our formalism to 
binary systems with comparable masses, in the hope that 
our conclusions based on the $\mu/M \to 0$ limit will be 
qualitatively valid also in the large-mass-ratio case. 
We note, in accordance with our previous observation, 
that in Eq.~(\ref{17}) $M$ is to be interpreted as
the total mass and $\mu$ as the reduced mass.

Our results are summarized in Table I. We consider
three types of binary systems. The first (system A) 
consists of two neutron stars, each with a mass equal to
$1.4 M_\odot$. The second (system B) consists of one 
neutron star ($1.4 M_\odot$) and one black hole 
($10 M_\odot$). The third (system C) consists of 
two black holes ($10 M_\odot$ each). 

The second column in Table I lists ${\cal R}_n$ as 
defined in Eqs.~(\ref{14}) and (\ref{17}), for
$n$ ranging from 4 to 8. This shows
that for system A, the signal-to-noise ratio is 
always smaller than 0.7651 times the
maximum possible value, with the best result 
obtained when $n=7$, which corresponds to $dE/dt$ 
accurate to 3.5PN order, and $dE/df$ accurate to 
3PN order. The reduction in signal-to-noise ratio
is therefore quite severe, even at such a high order
in the post-Newtonian expansion. The corresponding
results for systems B and C can also be obtained from
Table I. 

The reduction in signal-to-noise ratio is due to the fact 
that $P_n(v)$ is only an approximation to $P(v)$, 
see Fig.~1, and that $Q_n(v)$ is only an approximation 
to $Q(v)$. It is interesting to ask how much of the 
signal-to-noise ratio could be recovered if only $P(v)$ 
were approximated by a post-Newtonian expansion, while
$Q(v)$ were kept exact. To answer this question amounts to 
repeating the calculation presented previously, but with
$Q(v)$ substituted in place of $Q_n(v)$ in Eq.~(\ref{17}).
The results are shown in the third column of Table I. 
Not surprisingly, we see that keeping $dE/df$ exact gives 
much better results. For system A,
${\cal R}_n$ can become as large as 0.9864 (for $n=7$),
and is already as large as 0.9454 for $n=6$ (which corresponds
to $dE/dt$ accurate to 3PN order). The corresponding
results for systems B and C can also be obtained from 
Table I.

\begin{table}[t] 
\caption[Table I]{Reduction in signal-to-noise ratio 
incurred when matched filtering with approximate, post-Newtonian 
templates. For each of the considered binary systems, the
first column lists the order $n$ of the approximation,
the second column lists ${\cal R}^{(n)}$ as calculated using
the post-Newtonian approximation $Q^{(n)}$ for $dE/df$, the 
third column lists ${\cal R}^{(n)}$ as calculated using the
exact expression $Q(v)$ for $dE/df$, and the fourth column lists
${\cal R}^{(n)}$ as calculated using the alternative 
post-Newtonian approximation $Q^{\prime (n)}(v)$ for $dE/df$.}
\begin{tabular}{cccc}
$n$ & $Q^{(n)}(v)$ & $Q(v)$ & $Q^{\prime(n)}(v)$ \\
\hline
\multicolumn{4}{l}{System A ($1.4\ M_\odot\ + \ 1.4\ M_\odot$):} \\
\multicolumn{4}{l}{ } \\
4 & 0.5796 & 0.4958 & 0.4326 \\
5 & 0.4646 & 0.5286 & 0.6492 \\
6 & 0.7553 & 0.9454 & 0.9300 \\
7 & 0.7651 & 0.9864 & 0.9819 \\
8 & 0.7568 & 0.9695 & 0.9709 \\
\hline
\multicolumn{4}{l}{System B ($1.4\ M_\odot\ + \ 10\ M_\odot$):} \\
\multicolumn{4}{l}{ } \\
4 & 0.8478 & 0.3954 & 0.2916 \\
5 & 0.2413 & 0.2922 & 0.4012 \\
6 & 0.6097 & 0.6788 & 0.6051 \\
7 & 0.6023 & 0.9966 & 0.8528 \\
8 & 0.5734 & 0.8744 & 0.8942 \\
\hline
\multicolumn{4}{l}{System C ($10\ M_\odot\ + \ 10\ M_\odot$):} \\
\multicolumn{4}{l}{ } \\
4 & 0.8232 & 0.6919 & 0.4910 \\
5 & 0.3515 & 0.4422 & 0.5866 \\
6 & 0.7107 & 0.8088 & 0.7417 \\
7 & 0.7201 & 0.9997 & 0.9310 \\
8 & 0.6730 & 0.9152 & 0.9420
\end{tabular}
\end{table}

Why does the exact expression for $dE/df$ give such
better results? The answer is that while the exact
expression for $dE/df$ correctly vanishes at 
$v=v_{\rm isco}=6^{-1/2}$ (at the innermost stable 
circular orbit), its post-Newtonian analogue
fails to do so. For example, $Q_8(v)$ goes
to zero at $v\simeq 0.4236 > 6^{-1/2}$, corresponding
to a radius $r \simeq 5.572 M$.  

To establish that this indeed the 
reason, we consider an alternative
expression for $Q(v)$, obtained from Eq.~(\ref{4}) 
by expanding only the $(1-3v^2)^{-3/2}$ factor:
\begin{eqnarray}
Q'(v) &=& (1-6v^2) \Bigl( 1 +
{\textstyle \frac{9}{2}} v^2 +
{\textstyle \frac{135}{8}} v^4 
\nonumber \\ & & \mbox{} +
{\textstyle \frac{945}{16}} v^6 +
{\textstyle \frac{25515}{128}} v^8 +
\cdots \Bigr).
\label{18}
\end{eqnarray}
This expression, together with its truncated versions 
$Q'_n(v)$, manifestly go to zero at $v=6^{-1/2}$.
We have repeated the calculations with $Q'_n(v)$
substituted in place of $Q_n(v)$ in Eq.~(\ref{17}).
The results are shown in the fourth column of Table I. It
is evident that, generally speaking, using 
$Q'_n(v)$ gives much larger values of 
${\cal R}_n$ than using $Q_n(v)$. This is
especially remarkable for system A, since the 
relative contribution to the signal-to-noise
ratio coming from large values of $v$ is very small: 
For $v=6^{-1/2}$ and $M = 2.8 M_\odot$ the corresponding 
frequency is 1570 Hz, at which $S(f)$ is 200 times its
minimum value $S_0$.
 
These results suggest that accurate knowledge of the
location of the innermost stable circular orbit (where $dE/df$
goes to zero) might significantly improve the performance
of the post-Newtonian templates, through a factorization
of the kind shown in Eq.~(\ref{18}). It is possible that
in the case of binary systems with large mass ratios, 
such information could be obtained by numerically 
solving the initial value problem of general relativity
\cite{BlackburnDetweiler,Cook}, which would 
be far less laborious than solving the full 
dynamical problem. Another approach would be to use the 
Kidder-Will-Wiseman ``hybrid'' equations of motion 
\cite{KWW}, which give a better approximation to the
innermost stable circular orbit than the standard 
post-Newtonian equations. 

Conversations with Cliff Will were greatly appreciated.
This work was supported by the Natural Science Foundation
under Grant No.~PHY 92-22902 and the National Aeronautics
and Space Administration under Grant No.~NAGW 3874.


\begin{references}
\bibitem{Blanchet} The term of order $v^5$, complete with its
        $\mu/M$ dependence, was recently 
        calculated using post-Newtonian theory by Luc
        Blanchet (personal communication).
\end{references}
\end{document}


\draft
\title{Erratum and addendum: \\
       Gravitational radiation from a particle
       in circular orbit around a black hole. \\
       VI. Accuracy of the post-Newtonian expansion \\
       Phys.~Rev.~D 52, 5719 (1995)}
\author{Eric Poisson}
\address{Department of Physics, University of Guelph, 
         Guelph, Ontario, Canada N1G 2W1}
\date{\today}
\maketitle

A typographical error was discovered in the computer code
used to generate the results presented in the paper. The
corrected results are displayed in Table I, which also
incorporates new results, obtained using the recently 
improved post-Newtonian calculations of  Tanaka, Tagoshi, 
and Sasaki \cite{TanakaTagoshiSasaki}.

The quantity computed in Table I is
\[
{\cal R}_n \equiv \frac{S/N|_{\rm actual}}{S/N|_{\rm max}}
= \frac{ |(h|h_n)| }{ \sqrt{ (h|h) (h_n|h_n) } },
\]
the ratio of the actual signal-to-noise ratio obtained when
matched filtering a gravitational-wave signal $h$ with a
post-Newtonian template $h_n$ ($n$ denotes the order of
the approximation), to the maximum possible value. The
precise definition of the inner product $(\cdot | \cdot)$,
and all other technical details, are given in the paper. 

The reference signal $h$ is constructed from exact expressions
for both $dE/dt$, the rate at which energy is lost to gravitational 
waves, and $dE/df$, which provides the relation between orbital 
energy and gravitational-wave frequency. $dE/dt$ is obtained by 
numerically integrating the Teukolsky equation for a 
small-mass object in circular motion around a Schwarzschild
black hole. $dE/df$ is obtained by integrating the geodesic
equations for circular orbits. 

On the other hand, the template $h_n$ is constructed from 
post-Newtonian approximations to both $dE/dt$ and $dE/df$, accurate 
through order $v^n$, where $v = (\pi M f)^{1/3}$ (with $M$ the 
black-hole mass) is the orbital velocity. The expressions used 
are \cite{TanakaTagoshiSasaki}
\begin{eqnarray*}
\frac{dE/dt}{(dE/dt)_N} \equiv P(v) &=& 
1 - 3.7113 v^2 + 12.566 v^3 - 4.9285 v^4 - 38.293 v^5 + 
( 115.73 - 16.305 \ln v )v^6  - 101.51 v^7 \\
& & \mbox{} + 
( -117.50 + 52.743 \ln v )v^8 +  
( 719.13 - 204.89\ln v )v^9 + 
( -1216.9 + 116.64\ln v )v^{10} \\
& & \mbox{} +  
( 958.93 + 473.62\ln v )v^{11} + \cdots, \\
\frac{dE/df}{(dE/df)_N} \equiv Q(v) &=&
(1-6v^2)(1-3v^2)^{-3/2} \\
&=& 1 - {\textstyle \frac{3}{2}}v^2 - {\textstyle \frac{81}{8}}v^4
- {\textstyle \frac{675}{16}}v^6 - {\textstyle \frac{19845}{128}}v^8
- {\textstyle \frac{137781}{256}}v^{10} + \cdots \\
&=& (1-6v^2)\Bigl(1 + {\textstyle \frac{9}{2}}v^2 + 
{\textstyle \frac{135}{8}}v^4 + {\textstyle \frac{945}{16}}v^6 + 
{\textstyle \frac{25515}{128}}v^8 - 
{\textstyle \frac{168399}{256}}v^{10} + \cdots \Bigr),
\end{eqnarray*}
where $(dE/dt)_N$, $(dE/df)_N$ are the leading-order, Newtonian
expressions. The first form for $Q(v)$ is exact. The second form
(fully expanded) is a Taylor expansion about $v=0$. The third form
(partially expanded) forces $Q(v)$ to vanish at $v=6^{-1/2}$, the 
innermost stable circular orbit. The paper's original calculations
used post-Newtonian expansions accurate through $O(v^8)$. 

In calculating ${\cal R}_n$, the mass and arrival time parameters 
are assumed to be the same in $h$ and $h_n$. In the original
calculations, the initial-phase parameters were also assumed to
be matched. Here we proceed differently: the overlap integral
$(h|h_n)$ is {\it maximized} over all possible values of the
difference in initial-phase parameters. While this obviously 
leads to larger results for ${\cal R}_n$, the improvement is
not significant: at most 2\% when $n\geq 6$.

The paper's conclusions are drastically affected by the coding
error. 

First, the values obtained for ${\cal R}_n$ are now much larger 
than what was obtained before, and the Table shows that 
${\cal R}_n$ nicely converges to unity as the order of the 
post-Newtonian approximation increases. In particular, the Table 
shows that a post-Newtonian template accurate through order $v^6$ 
would reproduce more than 95\% of the signal-to-noise ratio. 
(System B is an exception.)

Second, there is no longer any evidence for the observation that 
an accurate knowledge of the location of the innermost stable
circular orbit would significantly improve the performance
of the post-Newtonian templates. Indeed, in most cases the 
numbers displayed in the fourth column of Table I, calculated 
using the partially expanded form for $Q(v)$, are not significantly 
larger than those displayed in the second column. (And in some 
cases they are significantly lower.)

Finally, it appears that the best results are obtained when using a 
$n=9$ template. In this case both $dE/dt$ and $dE/df$ are 
over-estimated by the post-Newtonian approximations. However, 
since $h_n$ depends on the ratio $df/dt = (dE/dt)/(dE/df)$, it 
appears that the individual discrepancies are best canceled with 
$O(v^9)$ expressions. 

The author wishes to apologize for any inconvenience caused
by his negligence. This work was supported by the Natural
Sciences and Engineering Research Council of Canada.

\narrowtext
\begin{table} 
\caption[Table I]{Reduction in signal-to-noise ratio 
incurred when matched filtering with approximate, post-Newtonian 
templates. For each of the considered binary systems, the
first column lists the order $n$ of the approximation,
the second column lists ${\cal R}_n$ as calculated using
the fully expanded expression for $Q(v)$, the third column 
lists ${\cal R}_n$ as calculated using the exact expression 
for $Q(v)$, and the fourth column lists ${\cal R}_n$ as calculated 
using the partially expanded expression for $Q(v)$.}
\begin{tabular}{cccc}
$n$ & $Q$ fully exp. & $Q$ exact & $Q$ part.~exp. \\
\hline
\multicolumn{4}{l}{System A ($1.4\ M_\odot\ + \ 1.4\ M_\odot$):} \\
\multicolumn{4}{l}{ } \\
 4 & 0.6756 & 0.5852 & 0.5217 \\
 5 & 0.5565 & 0.6117 & 0.7157 \\
 6 & 0.9578 & 0.9451 & 0.9320 \\
 7 & 0.9777 & 0.9866 & 0.9802 \\
 8 & 0.9731 & 0.9772 & 0.9782 \\
 9 & 0.9848 & 0.9854 & 0.9858 \\
10 & 0.9836 & 0.9837 & 0.9838 \\
11 & 0.9836 & 0.9837 & 0.9838 \\
\hline
\multicolumn{4}{l}{System B ($1.4\ M_\odot\ + \ 10\ M_\odot$):} \\
\multicolumn{4}{l}{ } \\
 4 & 0.8580 & 0.4890 & 0.3772 \\
 5 & 0.6718 & 0.3788 & 0.4305 \\
 6 & 0.8271 & 0.7455 & 0.6859 \\
 7 & 0.8938 & 0.9977 & 0.8796 \\
 8 & 0.8762 & 0.8945 & 0.9111 \\
 9 & 0.9962 & 0.9650 & 0.9459 \\
10 & 0.9897 & 0.9958 & 0.9984 \\
11 & 0.9924 & 0.9977 & 0.9995 \\
\hline
\multicolumn{4}{l}{System C ($10\ M_\odot\ + \ 10\ M_\odot$):} \\
\multicolumn{4}{l}{ } \\
 4 & 0.8367 & 0.7610 & 0.5855 \\
 5 & 0.4433 & 0.5354 & 0.6684 \\
 6 & 0.9769 & 0.8500 & 0.7988 \\
 7 & 0.9038 & 0.9997 & 0.9469 \\
 8 & 0.9006 & 0.9314 & 0.9541 \\
 9 & 0.9998 & 0.9850 & 0.9669 \\
10 & 0.9941 & 0.9983 & 0.9995 \\
11 & 0.9959 & 0.9992 & 0.9999 
\end{tabular}
\end{table}